\documentclass[10pt,letterpaper]{article}
\usepackage[top=0.85in,left=2.75in,footskip=0.75in]{geometry}

\usepackage[utf8x]{inputenc}

\usepackage{textcomp,marvosym}

\usepackage{fixltx2e}

\usepackage{amsmath,amssymb}

\usepackage{cite}

\usepackage{microtype}

\raggedright
\usepackage[aboveskip=1pt,labelfont=bf,labelsep=period,justification=raggedright,singlelinecheck=off]{caption}

\makeatletter
\renewcommand{\@biblabel}[1]{\quad#1.}
\makeatother

\date{}

\usepackage{lastpage,fancyhdr,graphicx}
\usepackage{epstopdf}
\fancyheadoffset[L]{2.25in}
\fancyfootoffset[L]{2.25in}
\begin{document} 

\vspace*{0.2in}

\begin{flushleft}
	{\Large
		\textbf\newline{Correlations Between Human Mobility and Social Interaction Reveal General Activity Patterns} 
	}
	\newline
	\\
    Anders Mollgaard\textsuperscript{1},
	Sune Lehmann\textsuperscript{2}, 
	Joachim Mathiesen\textsuperscript{1*}
	\\
	\bigskip
	\textbf{1} Niels Bohr Institute,  University of Copenhagen, 2100 Copenhagen, Denmark
	\\
	\textbf{2} Technical University of Denmark, 2800 Kgs.~Lyngby, Denmark.
	\\
	\bigskip
* mathies@nbi.dk
\end{flushleft}

\section*{Abstract}
A day in the life of a person involves a broad range of activities which are common across many people. Going beyond diurnal cycles, a central question is: to what extent do individuals act according to patterns shared across an entire population? Here we investigate the interplay between different activity types, namely communication, motion, and physical proximity by analyzing data collected from smartphones distributed among 638 individuals. We explore two central questions: Which underlying principles govern the formation of the activity patterns? Are the patterns specific to each individual or shared across the entire population? We find that statistics of the entire population allows us to successfully predict 71\% of the activity and 85\% of the inactivity involved in communication, mobility, and physical proximity. Surprisingly, individual level statistics only result in marginally better predictions, indicating that a majority of activity patterns are shared across {our sample population}. Finally, we predict short-term activity patterns using a generalized linear model, which suggests that a simple linear description might be sufficient to explain a wide range of actions, whether they be of social or of physical character.
    
\section*{Introduction}
In general, human behavior is inherently difficult to predict and not least to model \cite{watts2011everything}, but recent advances in data collection techniques \cite{gonzalez2008understanding, salathe2010high, stehle2011high, eagle2009inferring, stopczynski2014measuring} have made it increasingly feasible to build and test models of human behavior at both the level of individuals and for larger populations.
There is a rich literature on predicting human behavior.
Within the study of human mobility, the focus has been primaryly exploring indivdual patterns \cite{song2010limits, lu2013approaching, gallotti2013entropic, cuttone2016understanding, mollgaard2016dynamics}, while some authors have worked to understand the interplay between human mobility and social activity \cite{calabrese2011interplay, cho2011friendship, wang2011human, sekara2016fundamental}. 
Recently, group and overall patterns of human mobility has also been used for mobility prediction \cite{lorenzo2012predicting,calabrese2010human}
In the area of social networks, prediciton has focused on predicting link formation (`link prediction'), see e.g.~\cite{liben2007link, Lu20111150}, as well as predicting the dynamics of link activity \cite{ugander2012structural, leskovec2009meme, weng2013virality} in order to predict dynamics spreading processes.
A broad range of models also exist in the field of online attention, see \cite{crane2008robust, mathiesen2013excitable, matsubara2012rise}. {Mostly unexplored is the feasibility of predicting individual dynamics based on general
patterns (see \cite{calabrese2010human,lorenzo2012predicting} for exceptions).}

{While prediction is a common thread in all of the studies referenced above, the actual predictions made vary broadly. 
The sense in which we can predict virality of a meme \cite{weng2013virality} is quite different from how we think about prediction within mobility \cite{cuttone2016understanding}.}
Here we aim to take the broadest possible view of activity, since the concept may include communication, face-to-face interaction, working, sleeping, etc.
By studying the interplay among different human activity types, we obtain a rich description of human dynamics.
{As we focus on this interplay between different types of activity, we investigate a simple representation of activity as a binary variable `active' vs.~`non-active' for each activity.}
A characterization of general activity patterns can provide a first step towards a comprehensive bottom-up description of collective human phenomena.

\begin{figure}
	\centering
	\includegraphics[width=.99\textwidth]{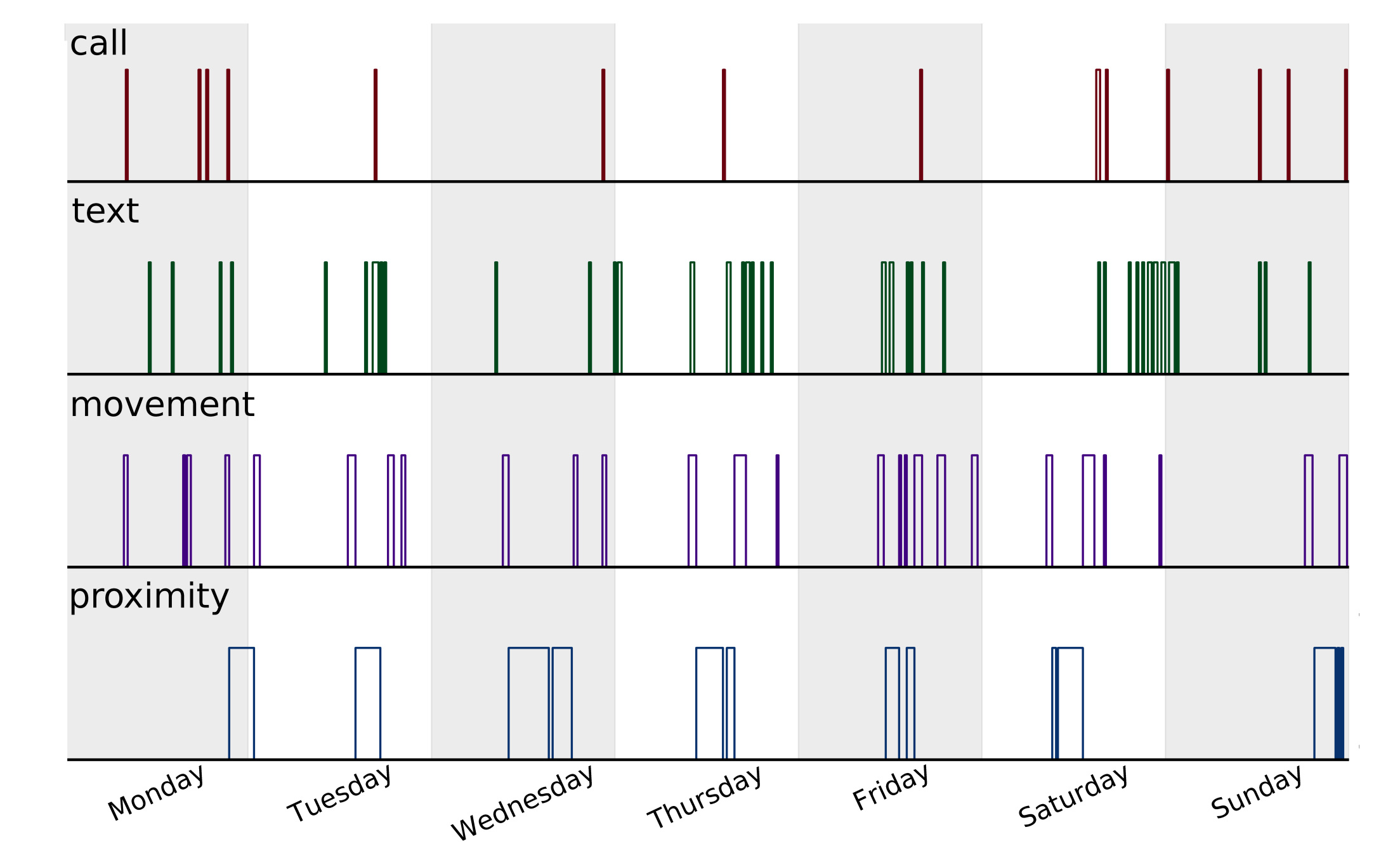}
	\caption{\textbf{Activity example.}We show the activity of a single user during one week of his life. From the top to the bottom panel we show the activity of calls (red), texts (green), movement (purple), and social proximity (blue). A daily pattern and a tendency for bursts can be noticed in the series, but further structure is not immediately visible. In the following we will discover the precise interactions between the different activities and use the emerging patterns to put upper bounds on the predictability of the future whereabouts of an individual .}
	\label{fig:circadian}
\end{figure}

We have analyzed data collected from smartphones distributed among 638 students at a large European University \cite{stopczynski2014measuring} (see Methods). Using custom software installed on the smartphones, we have gathered de-identified data regarding four types of activities, calling, texting (SMS), motion (from the GPS) and social proximity (via Bluetooth). {Here social proximity is defined as defined as physical closeness to another individual in the study (see Materials and Methods for details).} By slicing the data collection period of 18 months into bins of 15 minutes duration, we produce time series of the four activity types, $x^\mathrm{(u)}_i(t)\in \left\lbrace 0,1 \right\rbrace$, where $i$ denotes the type of activity for a given user, $\mathrm{u}$, at a time, $t$. Note that our representation is binary, such that each bin in the time series is represented by either activity or inactivity. In Fig.~\ref{fig:circadian}, we show an example of time series sampled for a single user during a given week. A daily pattern and a tendency for bursts \cite{wu2010evidence,vazquez2006modeling,zhou2008role,malmgren2008poissonian} is visible in the time series, but further structure is not immediately observable. In the following we seek to broaden our understanding of the underlying structure by answering the questions: How do the activities influence each other? Are the emerging patterns specific to an individual or general throughout the population? How well do the patterns predict the future?

\section*{Results}

\begin{figure*}[tbhp]
	\centering
	\includegraphics[width=.99\textwidth]{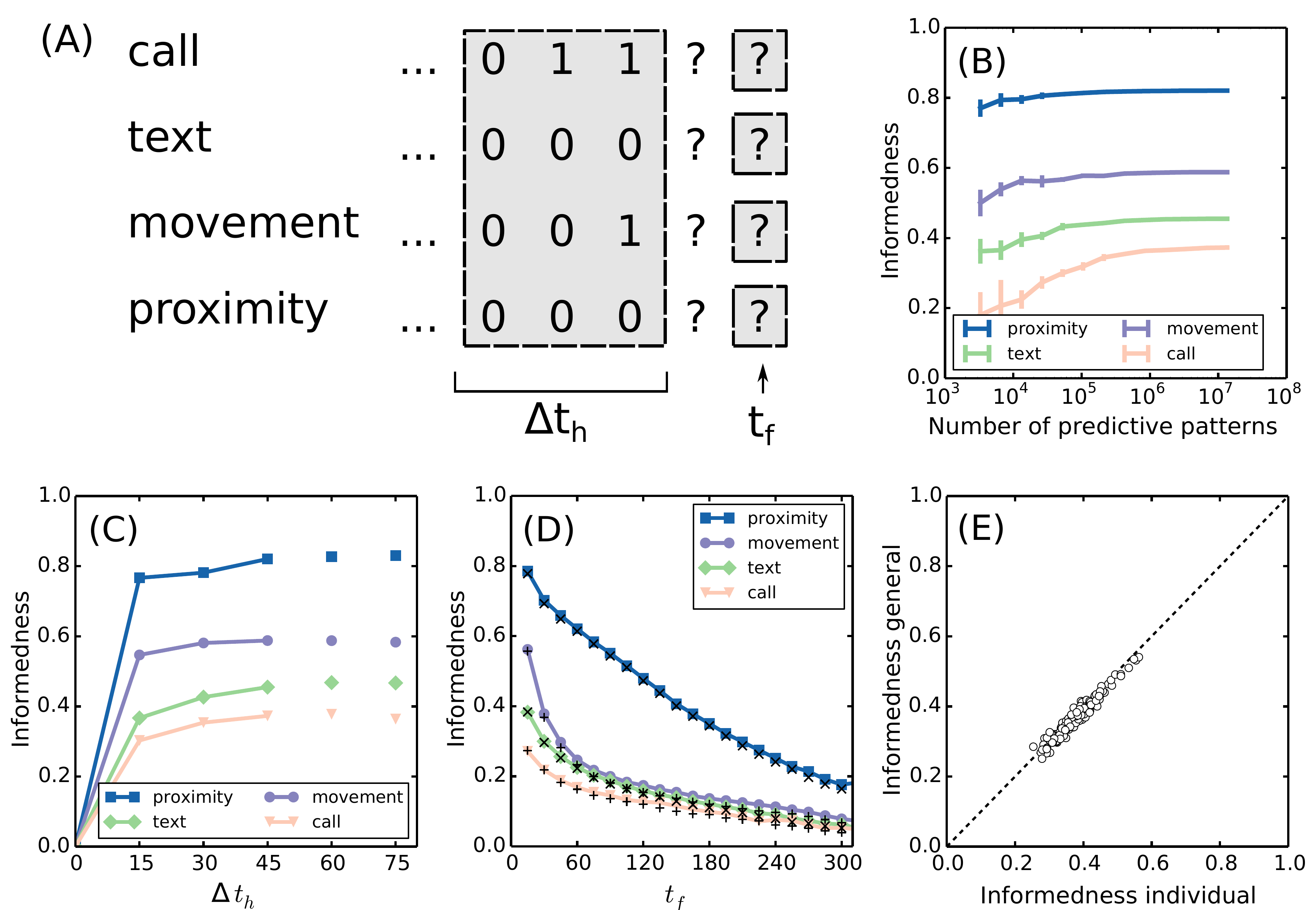}
	\caption{\textbf{Predicting future activity.} (A) We visualize the task of predicting future activity based on a four dimensional time series of length $\Delta t_h$. Each time bin has a width of 15 minutes and 0/1 represents activity/inactivity. The probability of activity at $t=t_f$ may be determined from the statistics of similar predictive patterns in the data set. For example, the particular predictive pattern shown here has a 45\% probability of activity of the mobility type at $t_f = 30 \, \text{min}$. (B) The data set needs a certain size in order to make accurate probability estimates. Here we show the informedness of the predictions at $t_f = 15 \, \text{min}$ based on a predictive pattern of length $\Delta t_h = 45\,\text{min}$ and varying data set sizes. Full convergence is obtained around $10^7$ predictive patterns, which means that the data set is sufficiently large for this task. (C) Using the full data set, we then make predictions for $t_f = 15 \, \text{min}$ and vary the length of the predictive pattern. By increasing the memory, we also increase the informedness of our predictions, which is clear from the connected markers. The disconnected markers at $\Delta t_h = 60\,\text{min}$ and $\Delta t_h = 75\,\text{min}$ are limited by statistics, meaning that the true upper bound on informedness is not obtained. (D) We then fix the length of the predictive patterns to $\Delta t_h = 45\,\text{min}$ and vary $t_f$. The lines represent the population average of the informedness for predictions that are based on individual activity patterns. The 'x' and '+' markers represent the population average of the informedness for predictions that are based on common activity patterns. The common patterns almost explain the full information of the individual patterns, which tells us that the activity patterns are general within our population. (E) We expand the population average in terms of individual data sets for the case of movement and $t_f = 30 \, \text{min}$ (second predictive pattern from left, purple line, previous figure). The horizontal axis shows the informedness of predictions based on individual predictive patterns, while the vertical axis shows the informedness of predictions based on general predictive patterns. Both measures vary across individuals, but they do so in almost perfect agreement. }\label{fig:matrix}
\end{figure*}

If human activity follows distinct patterns, then the future activity of an individual may be predicted with some certainty. Our first task is to quantify the degree of predictability inherent to our data set. {We define a predictive pattern (the predictor variable) to be a set of consecutive time bins of total length $\Delta t_h$, which is separated from a future activity state (the target variable) by a time $t_f$. In Fig.~\ref{fig:matrix}A we show an example of a predictive pattern for the case of  $\Delta t_h = 45\,\text{min}$ and  $t_f = 30\,\text{min}$. Here, the predictor variable is a matrix with four rows (the activities) and three columns (the time bins). The target variable consists of four activities in one time bin.} Person Z is currently on the move talking to someone on the phone and we know that another call was made 15 minutes ago, but apart from that there has been no other activity during the last 45 minutes. The task is to make the best possible prediction of the future state based on the observed history. 

Before moving on to the actual predictions, it is important to note that the data set has a strong class imbalance. In particular, calls are present in only 2.7\% of the time bins, texts in 7.3\%, movement in 9.4\%, and social proximity in 8.8\%. The fraction of correct predictions is therefore not a useful measure of predictability, since guessing ``no activity" in all bins will yield a very high score. We address this bias by instead measuring the \emph{informedness} \cite{powers2011evaluation} of our predictions, $I=R_{11} + R_{00} - 1$. Here $R_{11}$ and $R_{00}$ are the fractions of predictive patterns with respectively active and inactive future states that have been correctly predicted. Consistently predicting either state or guessing at random will result in an informedness of 0. For instance, by guessing "no activity" for all the predictive patterns we will succeed in predicting all future inactivity ($R_{00}=1$) and fail in predicting any future activity ($R_{11}=0$), thereby obtaining an overall informedness of 0. Note that the measure of informedness applies to an individual activity type, so for each $\Delta t_h$ and  $t_f$ there will be four different numbers of informedness.

{Now suppose that we have a model that can estimate the probability of being active in a future state. How do we turn this probability into a prediction such that the informedness is maximized? As shown in Materials and Methods, one can optimize the expected informedness by using the following decision rule:

\begin{align}
x_{i,\mathrm{predict}} = \begin{cases}
1, & \mathrm{if} \;\;  p_i > \frac{n_i}{N}.\\
0, & \text{otherwise}.\label{eq:decision rule}
\end{cases}
\end{align}

\noindent Here $x_{i,\mathrm{predict}}$ is the prediction for the future state (active/inactive), while $p_i$ is the probability of being active as estimated by the model. $N$ is the size of the data set and $n_i$ is the number of predictive patterns with future activity of type $i$.} The ratio $n_i/N$ is the probability of future activity of type $i$ for a random predictive pattern. The rule therefore tells us to predict the future state (active/inactive) that is favored by the model as compared to random. 

{In order to apply the above decision rule we need a model to provide us with $p_i$. We shall first use a non-parametric statistical model based on the counting statistics of all predictive patterns. For example let us estimate the probability of someone being on the move. If we train our expectations of a given target variable from all predictive patterns ($N\sim13.5$ million) sampled over the whole population, we have approximately $n_\textrm{movement}\sim1.2$ million targets sampled at $t_f = 30\text{min}$ with movement activity, i.e.\ a ratio of 9\%. However, if we focus solely on the predictive pattern presented in Fig.~\ref{fig:matrix}A, i.e. the pattern characterized by two recent calls and some movement, then the ratio is instead 45\%. These 45\% represents our best estimate for the probability of movement in the target variable (the future state) given the this particular predictive pattern.} According to the decision rule in equation~(\ref{eq:decision rule}) we optimize the informedness by anticipating movement at $t_f = 30\,\text{min}$, since $p_\textrm{movement}=0.45$ is greater than $n_\textrm{movement}/N=0.09$. Note that the full information of the predictive pattern is exploited in making this prediction, so it will not be possible to get a better prediction with any other model trained on the same data. The only limitation is the size of the data set, which must be large enough to estimate $p_i$ with sufficient precision. 

Let us start by showing that our data set is sufficiently large for the case of $\Delta t_h = 45\,\text{min}$ and $t_f = 15\,\text{min}$. For that purpose, we test the {informedness} found on a subset of the data. Within the subset, we repeatedly apply the decision rule equation (\ref{eq:decision rule}) to obtain predictions for all the future states and these can then be compared with the actual values to compute the informedness. In Fig. \ref{fig:matrix}B we show the dependence of the informedness on the size of the data set. We find that the informedness converges for all four activity types before the full data is used. However, for longer predictive patterns, we do not observe full convergence (see the Materials and Methods paragraph on the informedness convergence) indicating that $\Delta t_h = 45\,\text{min}$ is the longest predictive pattern for which we can compute the upper bound on the informedness. In Fig. \ref{fig:matrix}C, we show the dependence of the informedness on the length of the predictive pattern. The connected data points correspond to true upper bounds on the informedness, while the unconnected data points are limited by statistics. Here we use the full size of the data set and fix $t_f = 15\,\text{min}$. The graph clearly illustrates that the system has a memory, since the informedness improves as more of the past is included (when not limited by statistics). Predictive patterns of length  $\Delta t_h = 45\,\text{min}$ allow us to successfully predict 71\% of the activity and 85\% of the inactivity for the four activity types in the near future, $t_f = 15\,\text{min}$.

\subsubsection*{Individual patterns} In the analysis above, the predictive patterns from everyone are included in the statistics such that patterns of individuals are not visible. Therefore, the common data set might be seen as a general representation of a human being without distinct characteristics. We saw that the near future ($t_f = 15\,\text{min}$) of the general human could be predicted with a high precision. Now we will have a look at individual data sets and compare the predictions of personal patterns to the predictions of the general patterns. The data on a single individual is only a small fraction of the full data set, providing less data for reliably estimating the probability of a predictive pattern. In the following, we therefore restrict the analysis to persons with at least $3 \cdot 10^4$ predictive patterns, which leaves us with 139 individuals. {Note that this threshold is not enough to ensure optimal probability estimates (see Fig. \ref{fig:matrix}B), but it does leave us with the subset of individuals that are the richest in data and still of a substantial size.} For each individual data set, we now perform the same analysis as above for varying $t_f$, but with the length of the predictive pattern fixed to $\Delta t_h = 45\,\text{min}$. We then average over the informedness obtained for the individual data sets, see Fig.~\ref{fig:matrix}D. As expected, we find that the predictions are best for the near future. There is a sharp drop in the informedness over the first 60 minutes followed by a linear decay. 

Next we compare predictions made using general patterns with the performance of individual based predictions. To do this, we investigate the informedness of predictions based on common activity patterns. 
For each individual, we first sample a subset of predictive patterns from the common data set, which has the same statistics as the individual data set. In particular, we sample such that any predictive pattern occurs exactly one time less in the common data sample than in the individual sample, because this ensures similar statistics for the prediction step. Note that even though the statistics of the predictive patterns are identical, the statistics of the future states are not. By applying the decision rule equation (\ref{eq:decision rule}) to the sample, we might therefore get another set of predictions. The predictions of the common data are tested and the informedness is computed. The average informedness across all individuals is shown in Fig.~\ref{fig:matrix}D with ``x'' and ``+'' markers. Somewhat surprisingly, the predictions of the general patterns almost fully match the predictions based on individual patterns. This shows that individual characteristics do not not change the quality of our predictions - or to put it another way - activity patterns are general across the population. Note that the occurrence of the different patterns is not the same across individuals. Some individuals, for instance, make a lot more calls than others and therefore have more patterns with many calls. But when we see the same predictive pattern for two individuals, then the same future prediction optimizes our informedness. In Fig.~\ref{fig:matrix}E we plot, for each individual, the informedness of the individual patterns against the informedness of the general patterns for future mobility at $t_f = 30\,\text{min}$. This plot corresponds to the data used to calculate the second purple disc from the left in Fig.~\ref{fig:matrix}D, but without performing the population average. We see that the predictability varies a lot among the individual data sets, but the performance of the individual- and general predictive patterns scale together. 
Finally, we note that the informedness at $t_f = 15\,\text{min}$ is strongly correlated to the informedness at $t_f > 15\,\text{min}$. It drops from $C\approx 1$ to $C\approx 0.5$ on the time scale of one hour (calls) to several hours (texts, movement, proximity). Therefore, people who are more predictable on short time scales are also more predictable on long time scales.

\subsubsection*{Activity patterns are linear} 
\begin{figure*}
	\centering
	\includegraphics[width=11.4cm]{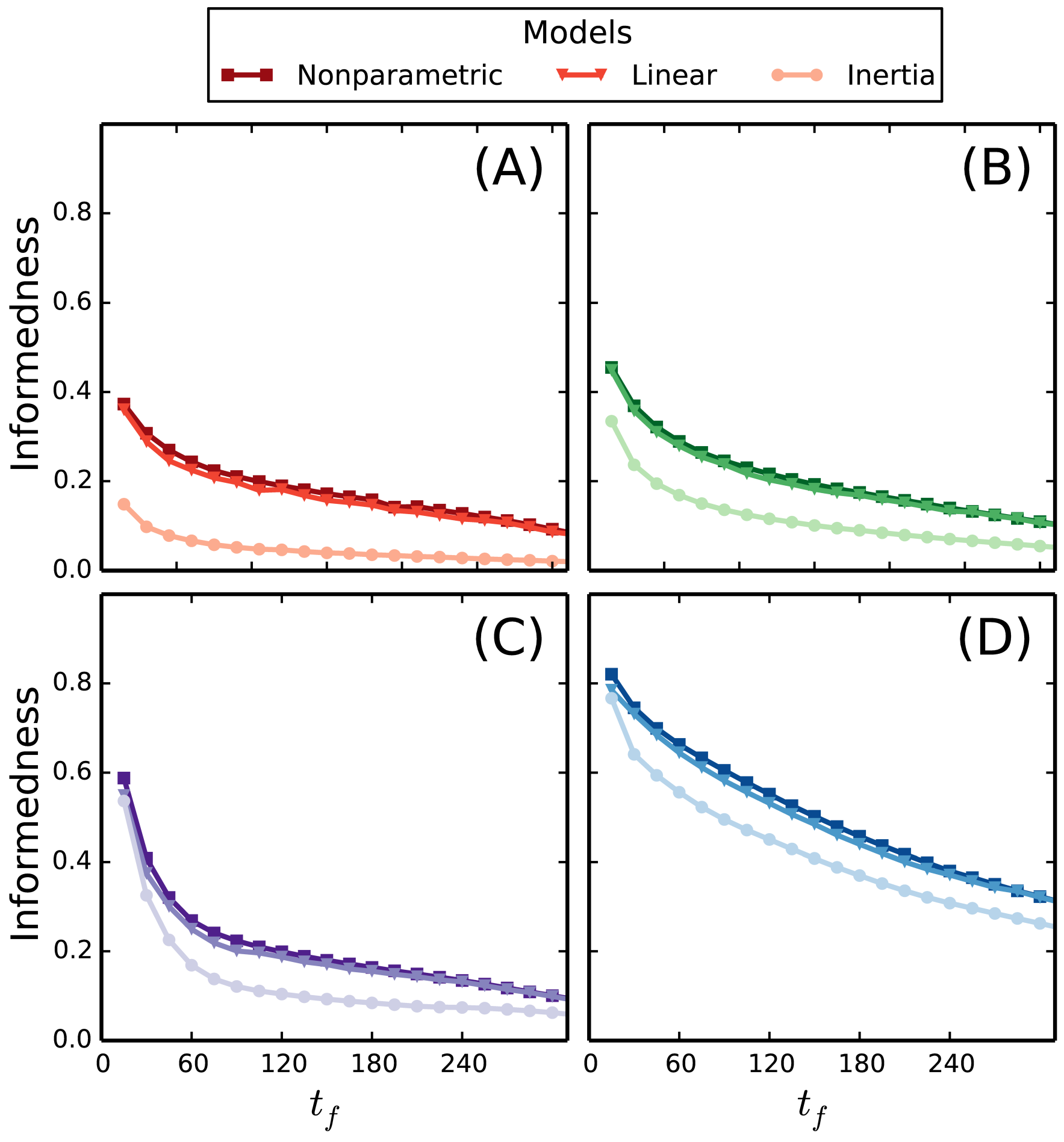}
	\caption{\textbf{Model comparison.} We show the informedness of our predictions regarding future activity of the respective activity types: call (A), text (B), movement (C), and proximity (D). The horizontal axis shows the reach of the predictions into the future in minutes. The informedness of predictions from three different models are presented. ``Nonparametric" (squares) is the label for the nonparametric model that was also presented in Fig.~\ref{fig:matrix}, here using a history length of $\Delta t_h = 45\,\text{min}$.  ``Inertia'' (discs) is a simple model which assumes that the future continues in the same state as the current one. We see that the near future predictions of movement and proximity is dominated by this information. ``Linear" (triangles) labels the predictions of a generalized linear model. Note that the informedness of the linear model almost matches the upper bound represented by the nonparametric model. }\label{fig:models}
\end{figure*}

In the previous section we exploited the size of the data set to perform separate statistics on all the possible predictive patterns. This nonparametric modeling approach is the optimal choice in terms of predictability, but it does not help us understand the underlying patterns. Furthermore, it is not possible to apply this method for very long predictive patterns and many activity types, since the number of possible predictive patterns grows exponentially with both. To incorporate a better time resolution, a longer history, and more activity types, one needs to apply a parametric approach. A parametric approach will necessarily make assumptions about the correlations in the data, which will never be perfectly accurate, but { it} allows us to train on much less data.  More importantly, we may learn about the central dynamics from the performance of simple models. Let us first consider a very simple model without any parameters; namely one which assumes that the future state is equal to the current one. We label this model ``Inertia'' and present the informedness of its predictions in Fig.\ \ref{fig:models} (discs). By comparing this result to that of the nonparametric model (squares), we conclude that the predictability of the near future for movement and proximity is mainly based on inertia. For predictions further into the future and for calls/texts, the dynamics is more complicated. We therefore introduce another model, which keeps things simple, but allows training to the history, namely the ``logit model''. The logit model is a generalized linear model, which feeds a weighted sum of the input to a sigmoid in order to find a probability estimate for the state of a binary system

\begin{align*}
p_i &= \sigma \left( \textbf{w}_i \cdot \textbf{x}_\text{history} \right), \\
&= \frac{1}{1+\exp \left( - \textbf{w}_i \cdot \textbf{x}_\text{history} \right) }.
\end{align*}

\noindent Here $\textbf{x}_\text{history}$ is a vector representing the activities and inactivities in a predictive pattern and $\textbf{w}_i$ is a weight vector, which is optimized within a training set according to L1 regularization \cite{scikit-learn}. We have used 75\% of the data set for training (chosen at random) and tested the model on the remaining 25\%. Note that the decision rule in equation (\ref{eq:decision rule}) again is applied to maximize the informedness, but now with $p_i$ determined from the linear model. The informedness of the predictions are shown in Fig.\ \ref{fig:models} (triangles). Somewhat surprisingly, the linear model almost matches the upper bound on informedness represented by the nonparametric model. A further analysis shows that the difference in informedness between the two models mainly arise from histories that are characterized by current inactivity, but with activity in the past. Here the non linear model tends to underestimate the probability of future activity; possibly because it would otherwise overestimate the probability of future activity for histories with activity both in the past and present. 

\begin{figure*}
	\centering
	\includegraphics[width=.99\textwidth]{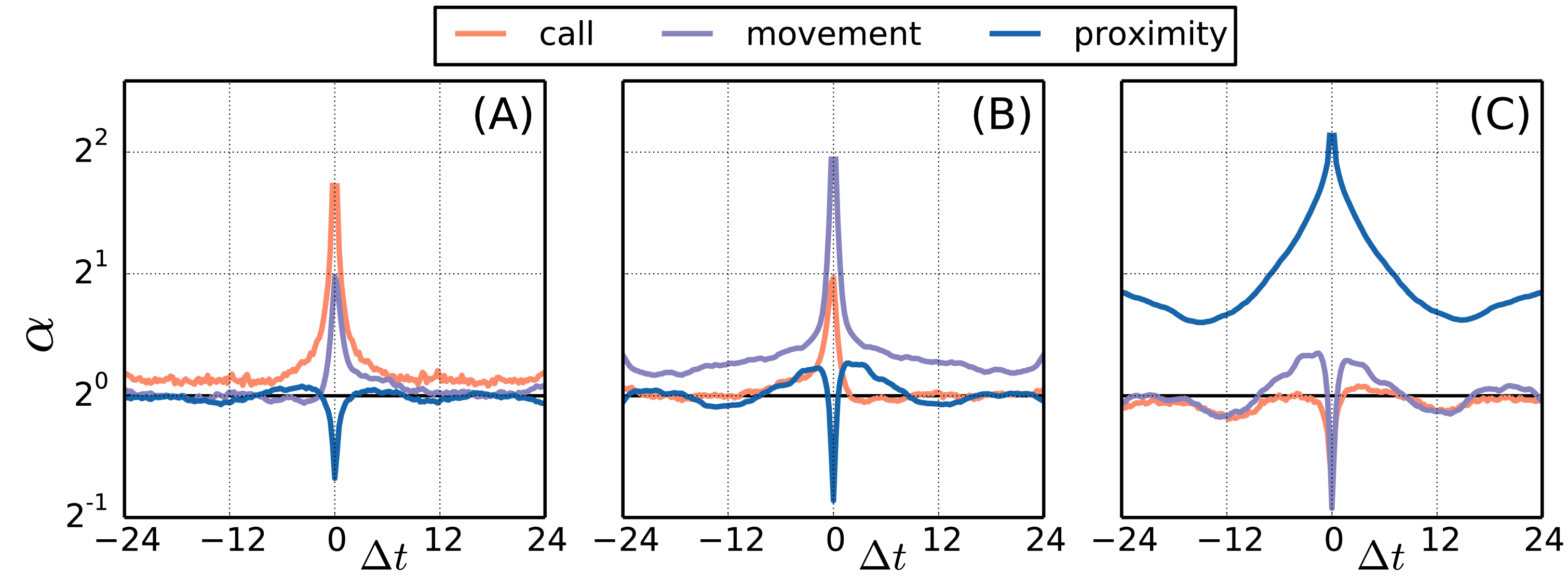}
	\caption{\textbf{Activity correlations.} In (A)-(C) we show the increase in activity triggered by respectively calling, movement, and social proximity. The horizontal axis spans 24 hours backward and forward in time and the vertical axis gives the factor of increased activity on a logarithmic scale. For example, the purple line in (A) tells us that movement is enhanced by a factor 2 at the time of a call and significantly increased for several hours after. More generally, all activity types are subject to self-enhancement effects, and significant cross-correlations are observed for all combinations of activities. }\label{fig:correlations}
\end{figure*}


\subsubsection*{Correlations} The success of the simple linear model reveals that the general dynamics foremost is driven by linear correlations. Note that the correlations are partly due to circadian patterns, since we tend to be active for all four activity types during daytime and inactive during night time. We stress that here we investigate correlations mediated by interactions, rather than by circadian cycles. First we compute the probability of an activity of type $j$ a time $\Delta t$ after observing an activity of type $i$, $P_{ij} (\Delta t)$. Since the behavior of this function is influenced by circadian cycles, we compute a reference probability function, $Q_{ij}(\Delta t)$ that contains the same circadian pattern, but which is unconditional on activity $i$ (see the Materials and Methods paragraph on Reference activity). The ratio of the two, $\alpha_{ij} (\Delta t) = P_{ij}(\Delta t) / Q_{ij} (\Delta t)$, measures the increase of activity $j$ conditioned on activity $i$ excluding the effects of time and weekday. In Fig. \ref{fig:correlations}A-\ref{fig:correlations}C we plot $\alpha_{ij} (\Delta t)$ for $i,j\in\ \left\lbrace \text{call, movement, proximity} \right\rbrace$. In Fig. \ref{fig:correlations}A we see the information gained by observing a call at time $\Delta t = 0$. We notice that the probability of another call 15 minutes after the first is increased by  a factor of 3.3; a self-enhancement that is visible across several hours. Similarly, we find that the probability of movement is increased by a factor of 2.0 in the vicinity of a call and with an asymmetric drop off that indicates an arrow of causality mostly pointing from call to movement. Finally, we find that the probability of social proximity is reduced by a factor 0.6 in the vicinity of a call and is slightly increased prior to a call. In \ref{fig:correlations}B we find that movement also shows self-enhancement, but with a peak value of 3.8 and a slightly slower decay. It is likewise found that the probability of social proximity is reduced by a factor 0.6 while moving, but that it is enhanced a few hours before and after moving. In \ref{fig:correlations}C we see that the self-enhancement of social proximity peaks at a value of 4.4 and decays on a time scale that surpasses 24 hours. This time scale is not representative of general social proximity, but is more likely to be an artefact of our measurements being restricted to a specific group of people that interact strongly in five day intervals (weekdays) and less so during weekends and holidays. The qualitative effects of texting (not shown) are very similar to calling. The general self-enhancement observed for all four activity types is in agreement with previous research stating that humans are characterized by long periods of inactivity followed by bursts of activity \cite{wu2010evidence,vazquez2006modeling,zhou2008role,malmgren2008poissonian}.


\section*{Discussion} We have introduced a framework for analyzing human activity patterns along with a decision rule that allows us to make optimal predictions in terms of informedness. We find that individuals to a large extent act according to a general set of activity patterns, thereby allowing us to predict individual activity with a high precision from general patterns. We believe that our conclusions based on four selected variables carries over to other variables related to human activity. Furthermore, we found that the optimal informedness of the full patterns were almost matched by a much simpler generalized linear model. We conclude that the self-enhancements and cross correlations presented in Fig. \ref{fig:correlations} provide an almost complete description of the general dynamics in the system to the extent of the available information. 

We should state that our analysis is subject to a number of limitations. First of all, we have considered only four activity types expressed as binary states. It would be interesting to include other variables such as sleeping, talking, online activity, etc. { Secondly, it might be possible that the homogeneity of our sample population makes the activity patterns more similar than in a fully randomly chosen population. However, in relation to our conclusion on the linear model, we do not find any reason to believe that university students have a more linear dynamic than the general background population. At the same time, we emphasize that making control studies across different demographics would require data very far beyond the already very comprehensive data material presented here.}
Finally, the true upper bounds on informedness has only been computed for resolutions of 15 minutes and history information below one hour. This restriction is necessary, since the number of predictive patterns needed to train the nonparametric model grows exponentially with time resolution, length of history, and number of variables. We therefore need to explore other approaches such as the generalized linear model, which almost match the nonparametric model in performance, while also allowing the extension to better time resolutions, longer histories, and more activity types.

\section*{Materials and Methods}
{The original data set involved 752 participants. We arrive at the 638 participants by filtering out individuals with less than 100 bins of activity among either of the activity types. The average number of active bins in the filtered data is 577 for calls, 1572 for texts, 2021 for movement, and 1894 for social proximity. The filtering is imposed to assure that the smartphone has been used as a main device. The data collection app and the data collection process is described in detail in \cite{stopczynski2014measuring}. The step from raw data to activity signals will be described below. This study was reviewed and approved by the appropriate Danish authority, the Danish Data Protection Agency (Reference number: 2012-41-0664). The Data Protection Agency guarantees that the project abides by Danish law and also considers potential ethical implications. All subjects in the study provided written informed consent.

}

 \subsubsection*{Data to signals}
We now briefly describe the methods that allow us to construct the activity time series from the data. The first step for all four activity types (call, texting, movement (GPS) and social proximity (Bluetooth) ) is to reduce the data to time stamps representing activity. For the call and texting, the activity timestamps simply correspond to the points in time where a call/text message was made or received. For GPS and Bluetooth, it is a little more complicated since least two measurements are needed to determine whether a person is moving or being in social proximity of another individual. 

All the phones record GPS information at regular intervals. Typically, the GPS information is stored on the phone in intervals of 1,2 or 3 minutes. To obtain activity time stamps, we start with the first recorded GPS time stamp, $gps_1$ and search for a second time stamp, $gps_2$, in the interval between 14 and 16 minutes later. If one or more GPS signals are found then we choose the one that is closest to 15 minutes. If a person has moved 200 meters or more between $gps_1$ and $gps_2$ then we say that the person is active, else we say that the person is inactive. The 200 meter requirement is enough to differentiate between actual movement and noise in the data, since the typical accuracy of the GPS is around 30 meters \cite{stopczynski2014measuring}. The time stamp is chosen as the center of the interval between $gps_1$ and $gps_2$. We then continue this procedure by looking 14 to 16 minutes further ahead, now with $gps_2\rightarrow gps_1$. If no GPS update is found, then we simply look further ahead until we reach an update, set this as $gps_1$ and continue the procedure.

The Bluetooth updates are recorded every 5 minutes. The conversion to a time signal follows the above procedure by starting with a first Bluetooth update, $bt_1$, and searching 15 minutes ahead in time for a second update, $bt_2$. For each update, we count the number of people in the study that are within a range of 3 meters. The phones are synchronized to activate their Bluetooth and scan for 30 seconds at the same time. {The distance to other people are computed from the strength of the Bluetooth signal, which, at the 3 meter threshold, has an accuracy of approximately 1 meter \cite{sekara2014strength}}. If person ``A" observes another person at the two points $bt_1$ and $bt_2$, then we say that person A is in a social proximity in the time interval between $bt_1$ and $bt_2$, otherwise we say that the person is socially inactive. The requirement of observing a person at two points in time prevents persons quickly passing by to be listed as social encounters. We furthermore require that people are within 3 meters from each other, that is, two persons being further apart but still visible on the Bluetooth scans are not considered as being in social proximity. We emphasize that from the signal itself, we cannot determine if actual social activity is taking place. We just determine that whether two people are within ``social proximity".

We construct time series by slicing time into 15 minute intervals or bins. The call/texting activity is considered active in a time bin, if the bin contains at least one activity time stamp. If not, it is considered inactive. The GPS/Bluetooth channel is considered active in a bin, if the bin contains at least one activity time stamp. If it instead contains an inactivity time stamp, then it is considered inactive. If a bin does not contain a time stamp from either GPS or Bluetooth, or if no call/SMS is registered within a 4 days range, then we label a bin as ``not trusted". All such bins are not included in the data analysis. The raw time series and an accompanying description are available in the Supporting Information files S1 Text and S1 Dataset.

\subsubsection*{The scoring system}
Here we introduce a system to quantify the predictive power of a model that makes predictions regarding the state of a two-state system, $x \in \left\lbrace 0,1 \right\rbrace$.  Predictions are typically quantified by the fraction of guesses that turn out to be right, but such a measure is very insensitive to information in cases, where one outcome is much more likely than the other. Imagine a model which is able to return the probability of a big earthquake in Tokyo on a daily basis. The probability of an earthquake may be $p=0.001$ on average, but the model is able to separate days into low risk, $p=0.0001$, and high risk, $p=0.3$. Such a model carries a lot of information, but according to the typical measure of predictive power it is no better than simply guessing ``no earthquake" on all days. It is our goal here to use a scoring system that is sensitive to such changes in probability.

Let us consider an ensemble of instances $x_i$ with $i=1,...,N$ that may be split into two groups according to their state. Let $n_0$ and $n_1$ be the number of instances in state 0 and state 1 respectively. Based on a model we make guesses regarding the value of an instance. A score, $s_{kl}$, is given for each instance based on the guess, $k \in \left\lbrace 0,1 \right\rbrace$, and the actual state, $l \in \left\lbrace 0,1 \right\rbrace$. The four score values, $\left\lbrace s_{00}, s_{11}, s_{10}, s_{01} \right\rbrace$, will be chosen such that neither of the two states are preferred on average, while correct/incorrect guesses for all signal values yield a total score, $S \equiv \sum_i s^{(i)}$, of $\pm 1$. These requirements are written down

\begin{align}
0 &= n_1 s_{11} + n_0 s_{10},\\
0 &= n_1 s_{01} + n_0 s_{00},\\
1 &= n_1 s_{11} + n_0 s_{00},\\
-1 &= n_1 s_{01} + n_0 s_{10},\\
\end{align}
and solved to yield

\begin{align}
s_{11} &= \frac{1}{n_1},\\
s_{10} &= - \frac{1}{n_0},\\ 
s_{01} &= 0,\\
s_{00} &= 0.
\end{align}
There is a symmetric solution under $0 \leftrightarrow 1$, but here we stick to the one presented. A model that correctly labels $m_{11}$ of the instances in state 1 and incorrectly labels $m_{10}$ of the instances in state 0 will obtain a total score of 

\begin{align}
S &= \sum_i s^{(i)},\\
&= \frac{m_{11}}{n_1} - \frac{m_{10}}{n_0},\\
&= R_{11} - R_{10},\label{eq:score1}\\
&= R_{11} + R_{00} - 1,\\
&= R_{00} - R_{01},\\
&= 1 - R_{10} - R_{01}.\label{eq:score2}
\end{align}
Here $R_{kl}$ is the fraction in state $l \in \left\lbrace 0,1\right\rbrace$ that are labeled as $k \in \left\lbrace 0,1 \right\rbrace$. $R_{11},R_{00},R_{10},$ and $R_{01}$ are known as respectively the sensitivity, specificity, false positive rate, and false negative rate. The total score is therefore seen to equal the ``informedness'', as it is defined in the general literature. We have used $1=R_{00}+R_{10}=R_{11}+R_{10}$, which shows that there are actually only two independent $R$'s and this allows us to interpret the total score as any of the four combinations in \ref{eq:score1} to \ref{eq:score2}.

\subsubsection*{Optimizing the informedness of a model}
We will now calculate the strategy that a probabilistic model must apply in order to optimize the expected informedness of its predictions. Let $q=n_1/N$ be the probability that a random instance is in state 1. If a model predicts a probability $p$ of a particular instance being in state $1$, then labeling the instance as $1$ yields an expected score of

\begin{align}
\left< s \right>_{k=1,l} &= \sum_l p_l s_{1l},\\
&= ps_{11} + (1-p)s_{10},\\
&= \frac{p}{n_1} - \frac{1-p}{n_0},\\
&= \frac{1}{N} \left( \frac{p}{q} - \frac{1-p}{1-q} \right),\\ 
&= \frac{1}{N} \left( \frac{ p(1-q) - (1-p)q }{q(1-q)} \right),\\ 
&= \frac{ p - q }{Nq(1-q)}. \label{eq: sum l}
\end{align}
The denominator is a positive number, so we find that the expected score of labelling the instance as $1$ is positive if $p>q$ and negative if $p<q$. The expected score from labelling the instance as $0$ is always 0

\begin{align}
\left< s \right>_{k=0,l} &= ps_{01} + (1-p)s_{00},\\
&= p \cdot 0 + (1-p) \cdot 0,\\
&= 0.
\end{align}
From these two results it is clear that in order to maximize the informedness, a model should label the instance according to the state that is promoted as compared to the average abundance, $q$. So if $p>q$, we guess $x=1$, and if $p<q$, we guess $x=0$. This is the procedure used for maximizing the informedness of the predictions made by the models in the paper.

\subsubsection*{Informedness convergence with data set size} 
\begin{figure}
	\centering
	\includegraphics[width=0.99\textwidth]{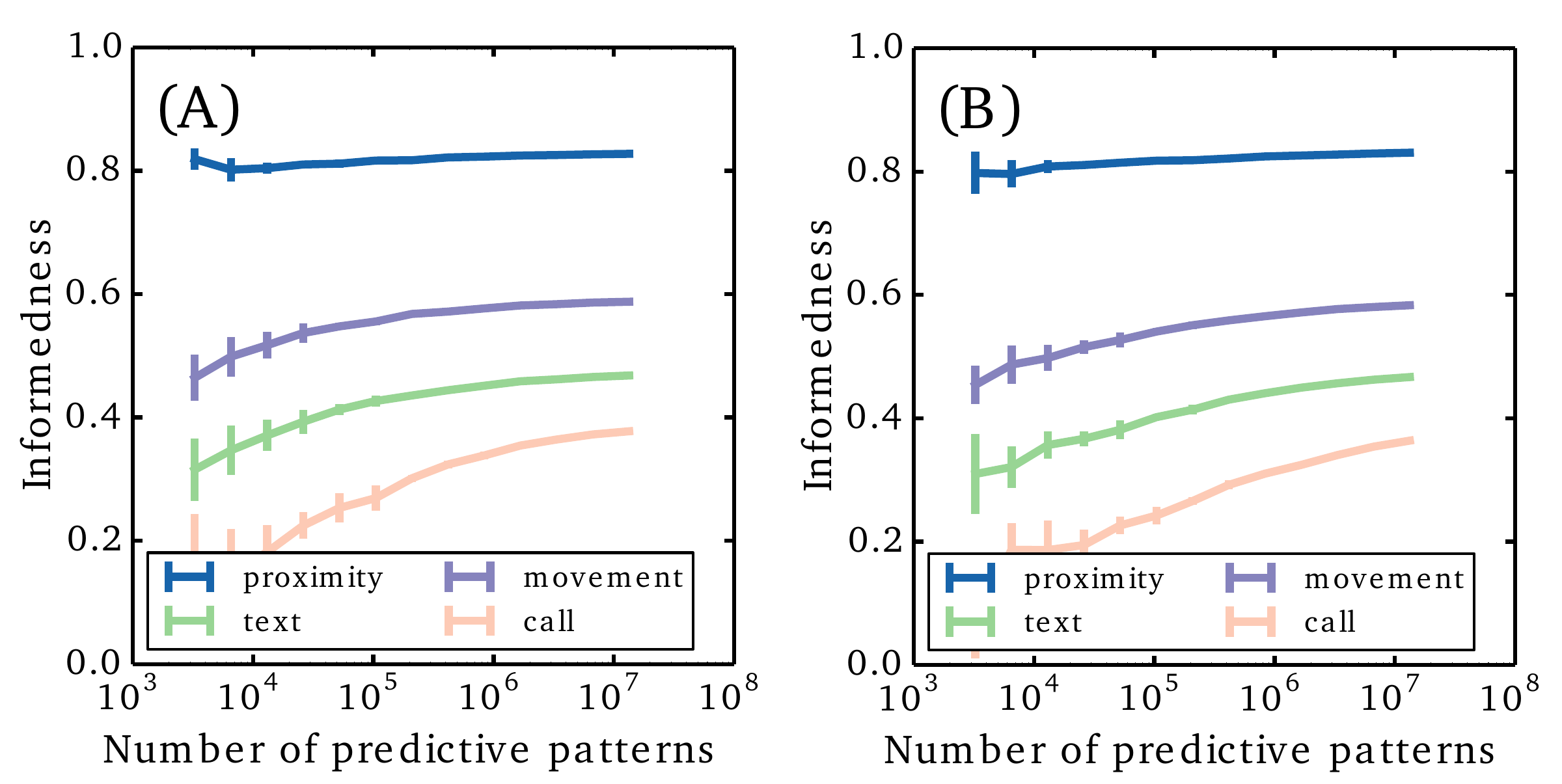}
	\caption{{\bf Convergence of informedness with the size of the data set.} We show the cases of 60 minute histories (A) and 75 minute histories (B). Note that full convergence is not reached within the full  size of the data set.}\label{fig:data size}
\end{figure}

The quality of the predictions regarding future activity depends on the size of the data set used for training. In the Results section, we show that our data set is large enough to obtain optimal predictions, when the history length is 45 minutes of duration. This is done by running the analysis on subsets of the full data set and check for convergence in informedness as the subset reaches the full data set size. In Fig.~\ref{fig:data size}, we show the same plots, but for the cases of 60 minutes and 75 minutes histories. Here we find that our data set is not large enough to reach full convergence; especially when predicting calls.

\subsubsection*{Reference activity}
The goal of this section is to explain the construction of the reference ensemble. The reference ensemble is introduced to quantize the increase in activity in a channel conditioned on activity in another channel, but without the effect of circadian patterns. Let $P_{ij}(\Delta t)$ be the probability of activity in channel $j$ conditioned on activity in channel $i$ with a delay $\Delta t$. It may be derived from the fraction of times that activity type $i$ is followed by activity type $j$ a time $\Delta t$ later. This quantity will however be heavily influenced by the diurnal cycle. To cancel the effect of diurnal cycles, we construct two ensembles representing the activity type $i$ for each user, $u$. The first ensemble simply contains the time stamps of all activity measured for the user $u$ of type $i$,  $E^\mathrm{(u)}_{1i} = \left\lbrace t \, | \, x^\mathrm{(u)}_i(t) = 1 \right\rbrace $. The second ensemble, $E^\mathrm{(u)}_{2i}$, is constructed by taking each time stamp in $E^\mathrm{(u)}_{1i}$, noting the weekday and time of day, and adding time stamps to $E^\mathrm{(u)}_{2i}$ corresponding to the same time of day at all equivalent weekdays during the full data collecting period. This way we get a reference ensemble with the exact same diurnal cycles, but with no correlation to the actual activity for times below a week, except through diurnal cycles.

\begin{figure*}
	\centering
	\includegraphics[width=0.9\textwidth]{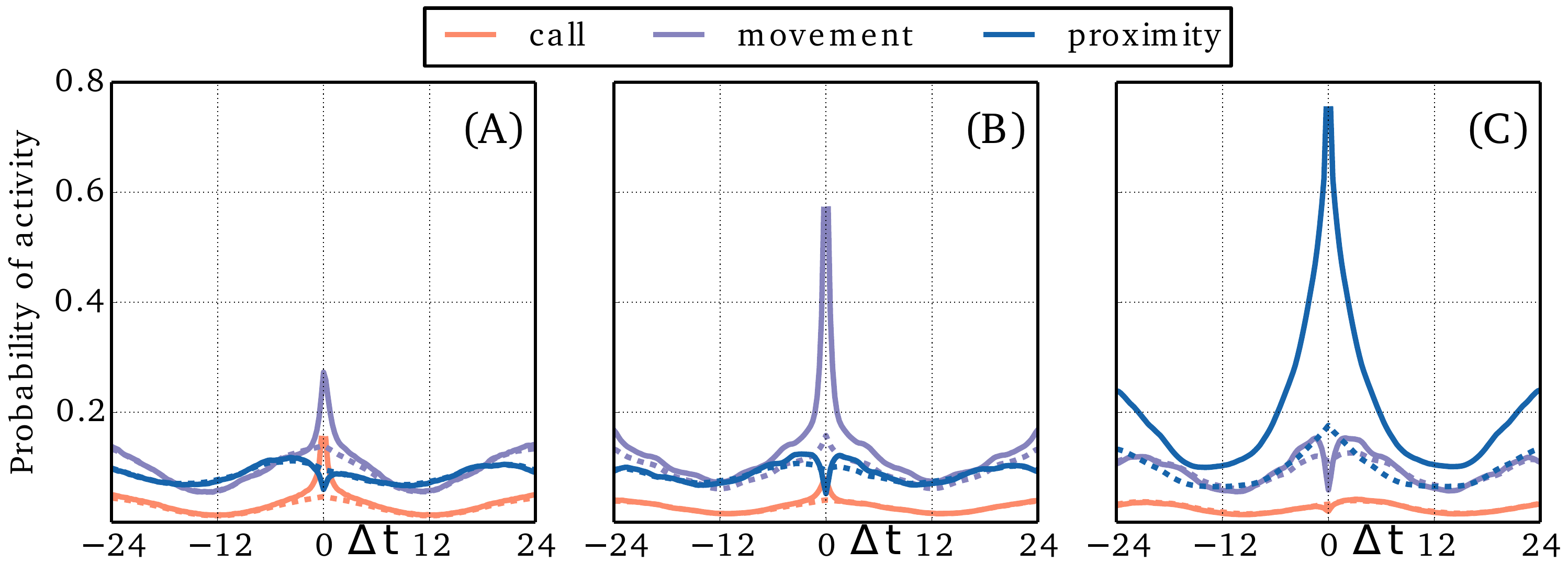}
	\caption{\textbf{Activity correlations.} In (A)-(C) we show the probability of activity conditioned on respectively a call, movement, and social proximity at $\Delta t = 0$. Also shown are reference activities (dashed), which takes into account the circadian correlations. The horizontal axis spans 24 hours backward and forward in time. }
	\label{fig:correlations2}
\end{figure*}

For each time stamp in $E^\mathrm{(u)}_{1i}$ we note the value of the signal in channel $j$, $x^\mathrm{(u)}_j(t)$, and the obtained count statistics allows us estimate the probability that user $u$ is active in channel $j$ given activity in $i$. The general probability, $P^\mathrm{(u)}_{ij} \left( \Delta t \right)$, that user $\mathrm{u}$ is active in channel $j$ a time $\Delta t$ after seeing activity in channel $i$ may be measured from the statistics of the delayed signal, $x^\mathrm{(u)}_j(t+\Delta t)$. We perform the same procedure for the time stamps in $E^\mathrm{(u)}_{2i}$ to obtain the probability, $Q^\mathrm{(u)}_{ij} \left( \Delta t \right)$, that the user would be active in channel $j$ regardless of any activity in channel $i$. For large delays, $\left| \Delta t \right| \gg \text{correlation length} $, we expect $P^\mathrm{(u)}_{ij} \left( \Delta t \right) = Q^\mathrm{(u)}_{ij} \left( \Delta t \right)$ and this is indeed what we find (see Fig. \ref{fig:correlations2}). The ratio between $P$ and $Q$ is shown in Fig. \ref{fig:correlations}.
\section*{Supporting Information}
{\bf S1 Text. Description of data format.} \\ \noindent
{\bf S1 Dataset. Dataset file}. The file is a gzip compressed text file in JSON format.

\end{document}